\documentclass[aps,prl,twocolumn,superscriptaddress]{revtex4-2}
\usepackage{amsmath,amssymb,amsfonts}
\usepackage{bm, latexsym}
\usepackage[dvipdfmx]{graphicx}
\usepackage{bbm,times}
\usepackage[normalem]{ulem}
\usepackage[usenames,dvipsnames]{color}
\usepackage[pdftex]{hyperref}
\usepackage{braket}

\begin{document}

\title{Localization properties in disordered quantum many-body dynamics \\ under continuous measurement}
\author{Kazuki Yamamoto}
\email{yamamoto@phys.titech.ac.jp}
\affiliation{Department of Physics, Tokyo Institute of Technology, Meguro, Tokyo 152-8551, Japan}
\affiliation{Department of Physics, Kyoto University, Kyoto 606-8502, Japan}
\author{Ryusuke Hamazaki}
\affiliation{Nonequilibrium Quantum Statistical Mechanics RIKEN Hakubi Research Team, RIKEN Cluster for Pioneering Research (CPR), RIKEN iTHEMS, Wako, Saitama 351-0198, Japan}

\date{\today}


\begin{abstract}
We study localization properties of continuously monitored dynamics and associated measurement-induced phase transitions in disordered quantum many-body systems on the basis of the quantum trajectory approach. By calculating the fidelity between random quantum trajectories, we demonstrate that the disorder and the measurement can lead to dynamical properties distinct from each other, although both have a power to suppress the entanglement spreading. In particular, in the large-disorder regime with weak measurement, we elucidate that the fidelity exhibits an anomalous power-law decay before saturating to the steady-state value. Furthermore, we propose a general method to access physical quantities for quantum trajectories in continuously monitored dynamics without resorting to postselection. It is demonstrated that this scheme drastically reduces the cost of experiments. Our results can be tested in ultracold atoms subject to continuous measurement and open the avenue to study dynamical properties of localization, which cannot be understood from the stationary properties of the entanglement entropy.
\end{abstract}

\maketitle

\textit{Introduction}.---Localization is an exotic phenomenon where a quantum state fails to spread over the entire Hilbert space. One notable mechanism of localization is the disorder, which prohibits the system from undergoing chaotic dynamics. For example, arbitrary weak disorder in noninteracting 1D or 2D systems leads to Anderson localization \cite{Anderson58, Abrahams79}, and many-body localization (MBL) transitions occur when the disorder strength exceeds a critical value in interacting systems \cite{Basko06, Prosen08, Pal10, Bardarson12, Luitz15, Serbyn15, Imbrie16, Imbrie16Stat, Geraedts16, Luitz17, Nandkishore15, Altman15, Abanin19, Smith16, Choi16}. Recently, MBL has broadened its research arena to open quantum systems \cite{Levi16, Fischer16, Medvedyeva16, Luschen17, Hamazaki19, Hamazaki22, Nieu17, Cai20}, where it has been demonstrated that unique nonequilibrium phenomena emerge, such as anomalous logarithmic growth of the von Neumann entropy with the scaling collapse normalized by the dissipation rate \cite{Levi16}.

Measurement is yet another mechanism that has recently attracted great interest to localize a quantum state in nonunitary quantum circuits \cite{Fisher18, Smith19, Skinner19, Yaodong19, Jian19, Fisher23rev, Tang20, Choi20, Bao20, Gullans20X, Gullans20L, Vijay20, Szyniszewski19, Turkeshi20, Zabalo20, Fan21, Ludwig21, Lunt21, Sang21, Block21, Cote22, Sierant22, Zabalo22L, Oshima23, Cai22, Weinstein22, Kelly22} and continuously monitored systems \cite{Cao19, Alberton21, Turkeshi21, Turkeshi22, Piccitto22, Fuji20, Goto20, Szyniszewski20, Jian21, Van21, Doggen22, Minato21, Muller21, Buchhold21}. Remarkably, novel quantum phenomena that have no counterpart in a closed system have been observed: A notable example is the measurement-induced phase transitions (MIPTs) \cite{Fisher18, Smith19, Skinner19, Yaodong19, Noel22, Koh22}, which are typically characterized by phase transitions from a volume law to an area law in the entanglement scaling of the stationary state. Interestingly, a continuously monitored MBL system can be conveyed to the area-law entanglement phase with an infinitesimal measurement strength \cite{Lunt20, Ippoliti21, Boorman22, Lu21, Zabalo22}. However, while both the disorder and the measurement localize the wave function and suppress the entanglement spreading, it is still not clear whether they exhibit the same localization properties \cite{Szyniszewski22}.

\begin{figure}[b]
\includegraphics[width=8.5cm]{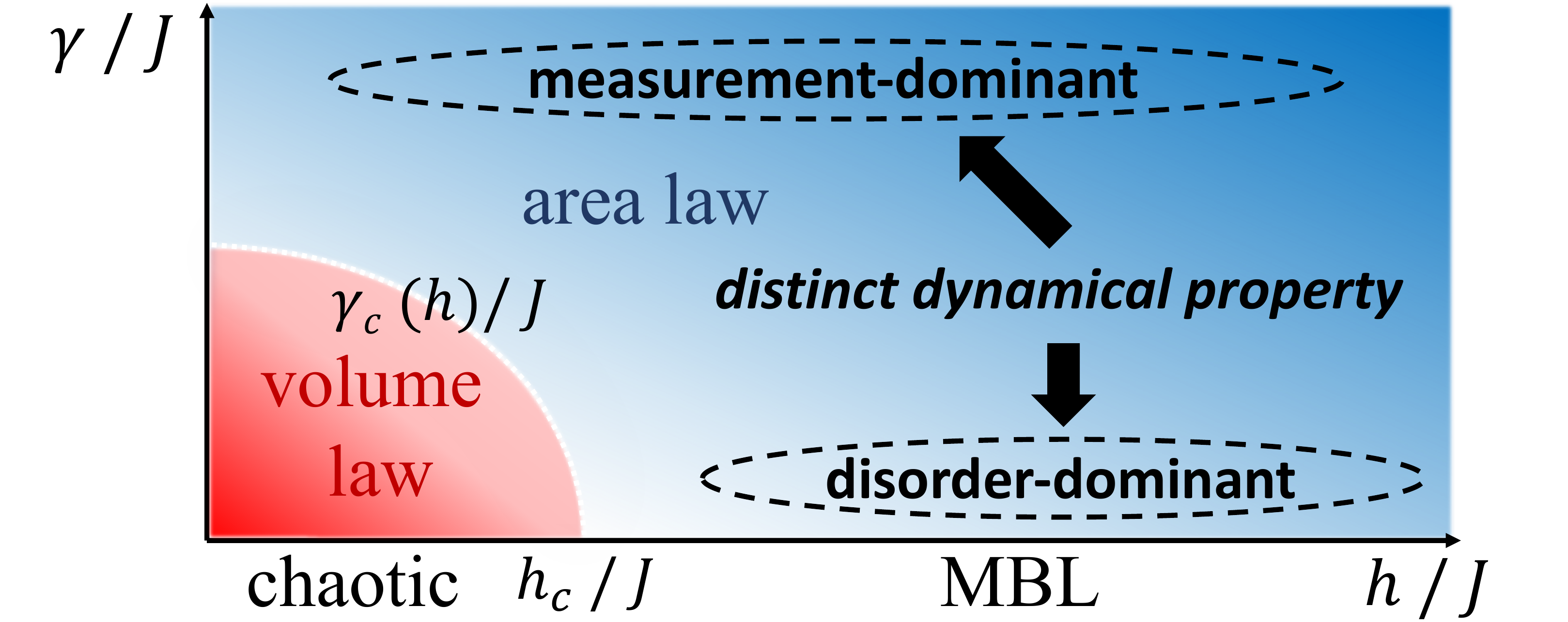}
\caption{Entanglement phase diagram with respect to the measurement rate $\gamma/J$ and the disorder strength $h/J$ in the steady state of the continuously monitored dynamics. Chaotic and MBL phases are separated at the critical disorder strength $h_c/J$ for $\gamma=0$, and MIPTs at the critical measurement rate $\gamma_c(h)/J$ only occur for $h<h_c$. Though two regimes surrounded by dashed black lines show the area-law entanglement, their dynamical properties of localization are distinct from each other, leading to the MMAL and the DMAL regimes.}
\label{fig_PhaseDiagram}
\end{figure}

In this Letter, we demonstrate that \textit{dynamical} properties of localization induced by the disorder and the measurement are distinct from each other in disordered quantum many-body systems under continuous monitoring. We first elucidate the whole entanglement phase diagram and associated MIPTs with respect to the disorder and the measurement strength. Then, by analyzing the fidelity between random quantum trajectories, we find that two dynamically distinct regimes in the area-law phase appear: The disorder-dominant measurement-induced area-law (DMAL) regime and the measurement-dominant measurement-induced area-law (MMAL) regime (see Fig.~\ref{fig_PhaseDiagram}). We show that the DMAL regime is characterized by an anomalous power-law decay of the fidelity. This distinction is further supported by the long-time dynamics of autocorrelation functions that relax to a disorder-independent value in the MMAL regime, while the DMAL regime exhibits slow dynamics reflecting the initial-state information.

Furthermore, to verify our result in an experimentally accessible way, we propose a general method to obtain physical quantities in the continuously monitored dynamics without postselection. We elucidate that, once jump processes are observed for a single trajectory, we can reproduce the dynamics by repeating the processes of Hamiltonian evolutions and appropriate unitary operations in a closed quantum system. This scheme significantly reduces the experimental cost compared to the postselection of the continuously monitored dynamics.

\begin{figure}[t]
\includegraphics[width=8.5cm]{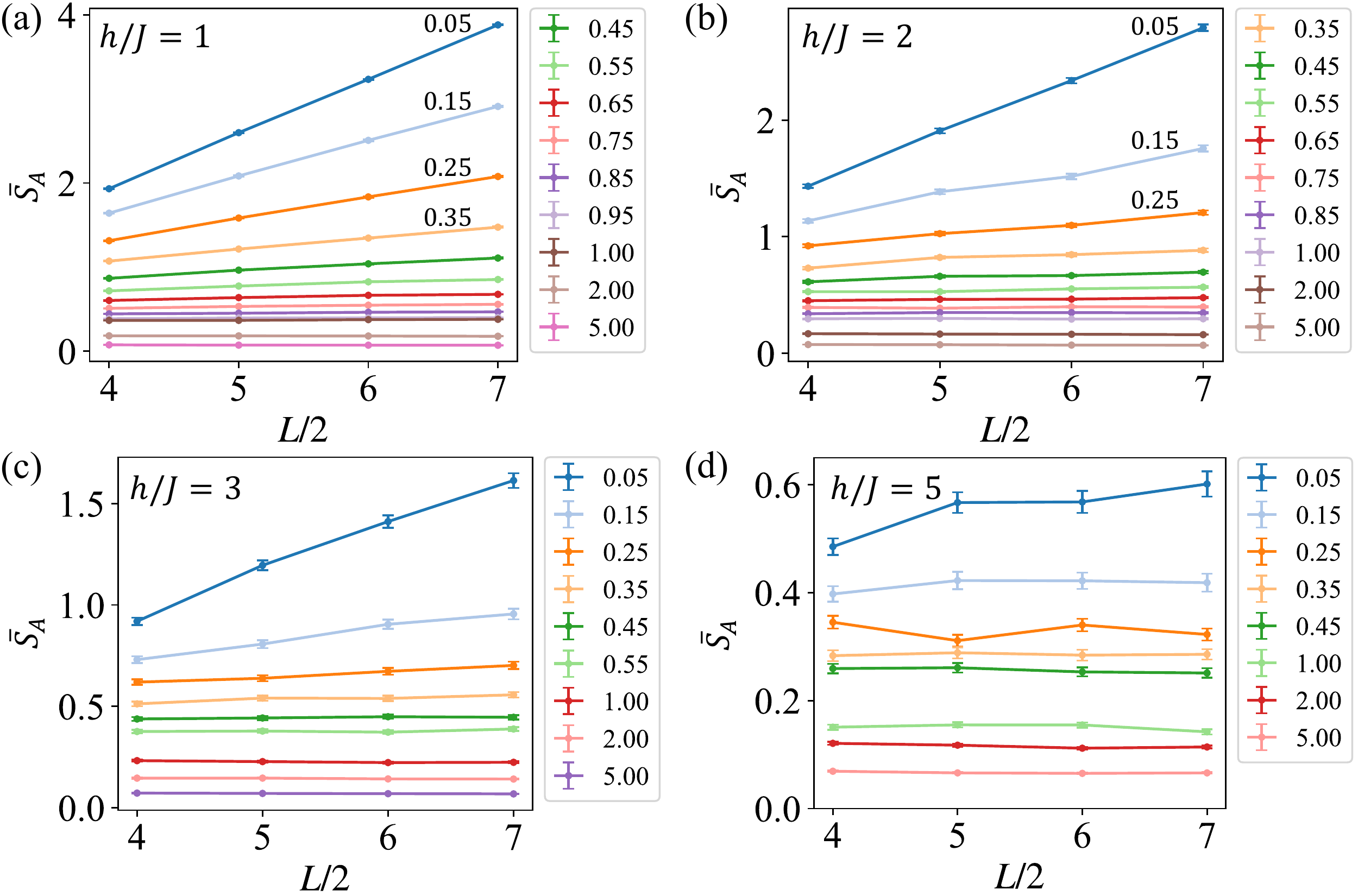}
\caption{Half-chain entanglement entropy $\bar S_A$ ($A=\{j|1\le j\le L/2\}$) in the steady state with respect to $L/2$ for (a) $h/J=1$, (b) $h/J=2$, (c) $h/J=3$, and (d) $h/J=5$. The legend shows $\gamma/J$. The entanglement exhibits MIPTs for $h<h_c$, while it immediately undergoes area-law transitions for $h>h_c$. Steady state is reached at $\gamma t=1000$ in (a), (b), and (c), and $\gamma t=2000$ in (d) with the average over $300\times 100$ realizations. Hereafter, we use the notation ``$n\times m$ realizations'' to represent that $m$-trajectory samplings are performed for each of $n$-disorder samplings. }
\label{fig_EntanglementEntropy}
\end{figure}

\textit{Measurement-induced phase transitions}.---We consider disordered interacting hard-core bosons on a one-dimensional lattice subject to open boundary conditions:
\begin{align}
H=\sum_{j=1}^{L-1}\frac{J}{2}(b_{j+1}^\dag b_j + b_{j+1}b_j^\dag) + \sum_{j=1}^{L-1}Vn_{j+1}n_j + \sum_{j=1}^L h_j n_j.
\label{eq_HB}
\end{align}
Here, $b_j$ $(b_j^\dag)$ is the bosonic annihilation (creation) operator satisfying the hard-core constraint $b_j^2=0$, $n_j=b_j^\dag b_j$ is the particle number operator, and disorder $h_j$ is randomly chosen from the uniform distribution $h_j\in[-h, h]$. Throughout this Letter, we set $V=J$. Since our aim is to study the measurement-induced trajectory dynamics in open quantum systems, we employ the stochastic Schr\"odinger equation obeying the marked point process \cite{Daley14, Fuji20}, which is given in the time interval $[t, t+dt]$ by
\begin{align}
d|\psi(t)\rangle=\big(&-iH_\mathrm{eff}+\frac{\gamma}{2}\sum_{j=1}^L\langle L_j^\dag L_j\rangle\big)|\psi(t)\rangle dt \notag\\
&+\sum_{j=1}^L \bigg(\frac{L_j|\psi(t)\rangle}{\sqrt{\langle L_j^\dag L_j\rangle}}-|\psi(t)\rangle \bigg)dW_j.
\label{eq_stochastic}
\end{align}
Here, $\langle\cdots\rangle$ denotes an expectation value with respect to the quantum state $|\psi(t)\rangle$, and $H_\mathrm{eff}=H-\frac{i\gamma}{2}\sum_j L_j^\dag L_j$ is the non-Hermitian Hamiltonian with the jump operator $L_j=n_j$, which describes continuous measurements of the local particle number. We note that a discrete random variable $dW_j$ satisfies $dW_jdW_k=\delta_{jk}dW_j$ and $E[dW_j]=\gamma\langle L_j^\dag L_j\rangle dt$, where $E[\cdot]$ represents an ensemble average over the stochastic process. Importantly, as the total particle number is a conserved quantity, the nonunitary dynamics under $H_\mathrm{eff}$ is replaced by the unitary evolution under $H$ upon the normalization of the state $|\psi(t)\rangle$ \cite{Fuji20}. In the following, we calculate the exact time evolution assuming that the initial state $|\psi_0\rangle$ is prepared in the N\'eel state $|1010\cdots\rangle$ at half filling. 

\begin{figure}[t]
\includegraphics[width=8.5cm]{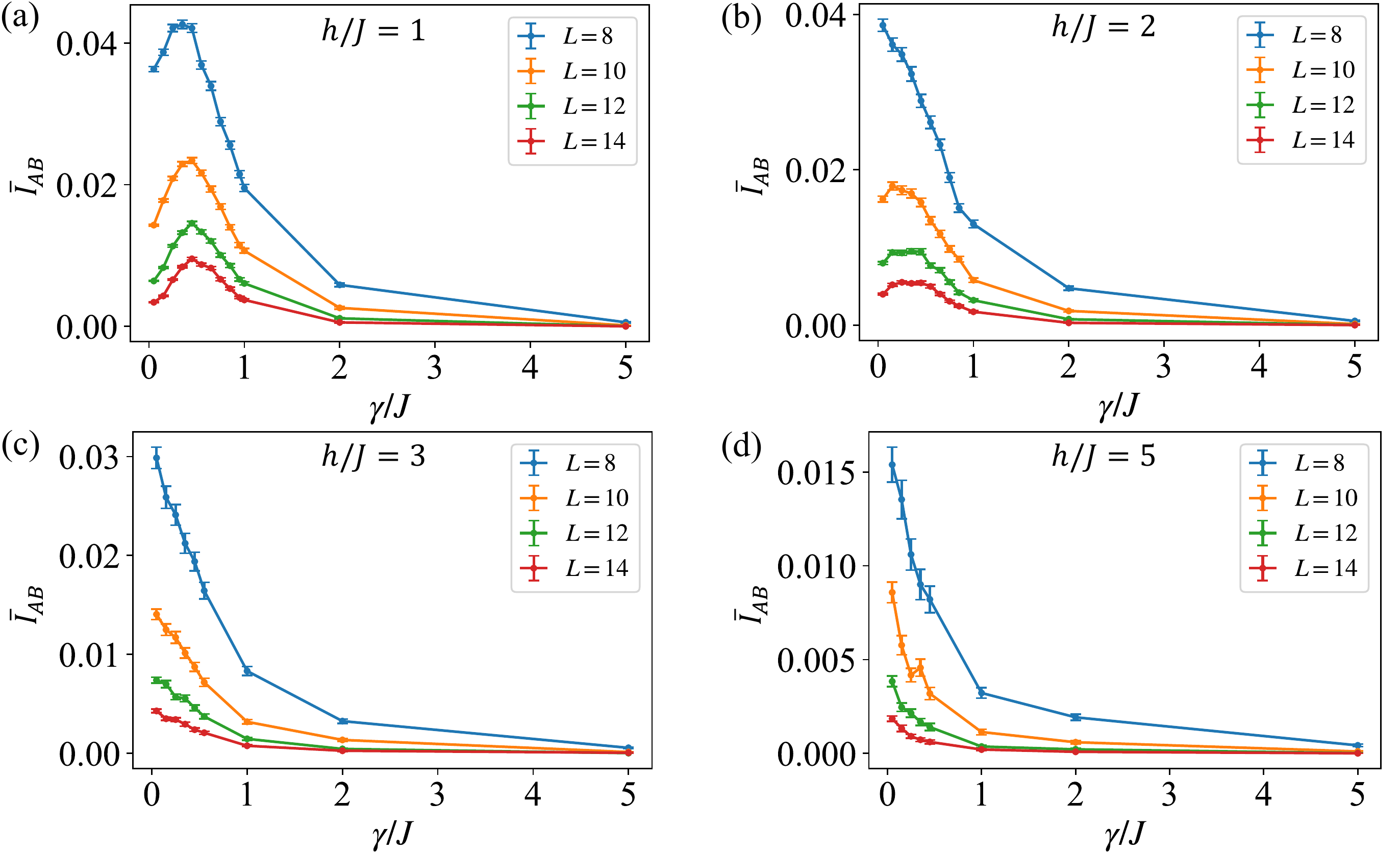}
\caption{Mutual information $\bar I_{AB}$ ($A=\{1\}$, $B=\{L/2+1\}$) with respect to $\gamma/J$ in the steady state for (a) $h/J=1$, (b) $h/J=2$, (c) $h/J=3$, and (d) $h/J=5$. Critical points of MIPTs are evaluated from the peak of $\bar I_{AB}$. Steady state is reached at $\gamma t=1000$ in (a), (b), and (c), and $\gamma t=2000$ in (d) with the average over $300\times 100$ realizations.}
\label{fig_MutualInformation}
\end{figure}

Under the quantum trajectory dynamics, the entanglement is built due to the Hamiltonian evolution, while it is erased by the measurement. To obtain entanglement properties and associated MIPTs in the steady state, we start by calculating the von-Neumann entropy for a subsystem $A$, which is defined by
\begin{align}
S_A=-\mathrm{Tr}_A[\rho_A\ln\rho_A].
\label{eq_EntanglementEntropy}
\end{align}
Here, $\rho=|\psi\rangle\langle\psi|$ is the density matrix of the system, and the reduced density matrix $\rho$ for a subsystem $A$ is obtained by tracing out over the complement $A^c$ as $\rho_A=\mathrm{Tr}_{A^c}\rho$. Figure~\ref{fig_EntanglementEntropy} shows the half-chain entanglement entropy $\bar S_A$ in the steady state with respect to subsystem sizes. Here, $\bar X=E_{\mathrm{dis}}[E[X]]$ with $E_\mathrm{dis}$ denoting the disorder average. We note that the entanglement entropy shows sufficient convergence with respect to time, and all physical quantities in the steady state obtained in this paper converge with respect to time, trajectory realizations, and disorder realizations with sufficient accuracy. From Figs.~\ref{fig_EntanglementEntropy}(a), (b), and (c), we see that the entanglement exhibits volume-law to area-law MIPTs as the measurement strength is increased in the chaotic phase with weak disorder. On the other hand, in the MBL phase with strong disorder, which is expected to emerge above the critical value $h_c/J\simeq3.6$ \cite{Pal10, Luitz15, Footnote}, it has been discussed that any finite measurements force the system to undergo area-law entanglement transitions \cite{Lunt20}. We indeed see that the results in Fig.~\ref{fig_EntanglementEntropy}(d) show the area-law entanglement scaling qualitatively well. We note that, in Ref.~\cite{Lunt20}, area-law transitions of entanglement
entropy were obtained only in the deep MBL phase, which is far above the critical point given by $h_c$.

To further quantify MIPTs, we calculate the bipartite mutual information $I_{AB}$ between two subregions $A$ and $B$ given by
\begin{align}
I_{AB}=S_{A}+ S_{B}- S_{{A}\cup{B}}.
\label{eq_MutualInformation}
\end{align}
Previous studies have found that the mutual information \eqref{eq_MutualInformation} exhibits a peak at the critical point of MIPTs with respect to the measurement rate \cite{Yaodong19}. Figure \ref{fig_MutualInformation} shows the mutual information in the steady state. We note that the choice of the distance between $A$ and $B$ does not affect the qualitative argument if they are located at a distance of the order of $L$; here, we take $A=\{1\}$ and $B=\{L/2+1\}$. As clearly seen from Figs.~\ref{fig_MutualInformation}(a) to (d), the peak of $\bar I_{AB}$ gradually shifts towards lower $\gamma/J$ as the disorder strength $h/J$ is increased, and in the MBL phase, the critical value of MIPTs $\gamma_c(h)/J$ becomes zero irrespective of the disorder strength. From these results, we have elucidated the phase diagram shown in Fig.~\ref{fig_PhaseDiagram} (see Supplemental Material \cite{Supple} for the quantitative diagram). We note that previous works on MIPTs in MBL systems have not obtained the entire phase diagram in Fig.~\ref{fig_PhaseDiagram} \cite{Lunt20, Ippoliti21, Boorman22, Lu21, Zabalo22}.

\begin{figure}[t]
\includegraphics[width=8.5cm]{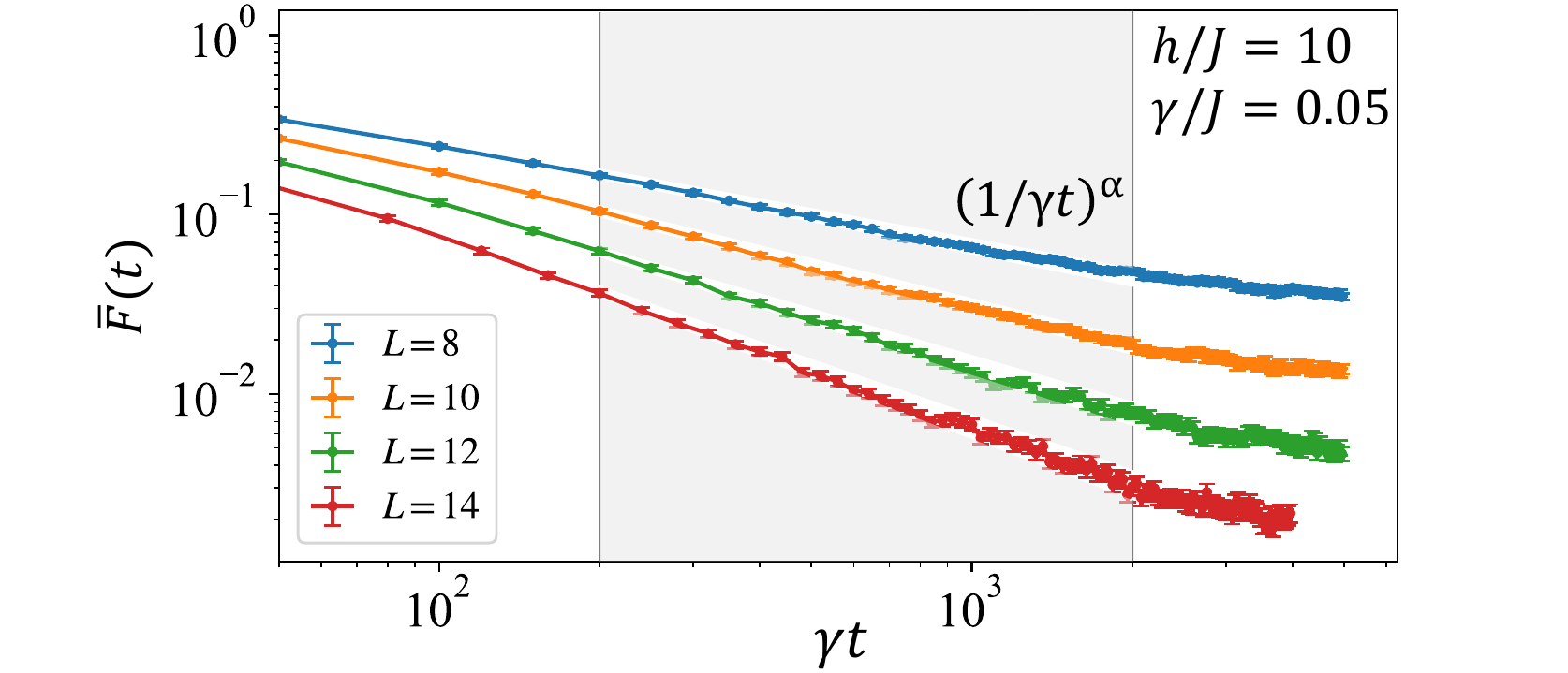}
\caption{Log-log plot of the dynamics of the fidelity $\bar F(t)$ with respect to $\gamma t$ in the DMAL regime. Shaded region, which is a guide to the eye, shows the power-law behavior of the fidelity. The parameters are set to $h/J=10$ and $\gamma/J=0.05$, and the disorder average is taken over $300$ realizations. For each fixed disorder, we take $50$ pairs of trajectories with respect to $|\psi(t)\rangle$ and $|\psi^\prime(t)\rangle$.}
\label{fig_Fidelity}
\end{figure}

\textit{Dynamical properties of localization}.---
Although the stationary state displays the area-law entanglement in the greater part of the phase diagram, it is not clear whether the regime near the unitary MBL phase (DMAL regime) and the one far from the unitary phase (MMAL regime) dynamically show the same property (see Fig.~\ref{fig_PhaseDiagram}). In order to clarify this problem, we analyze the fidelity that describes the overlap between independent random quantum trajectories $|\psi(t)\rangle$ and $|\psi^\prime(t)\rangle$ for each disorder given by
\begin{align}
F(t)=|\langle\psi(t)|\psi^\prime(t)\rangle|,
\label{eq_fidelity}
\end{align}
and study its average over disorder and trajectories \cite{Karol05}. We note that, as the (normalized) states $|\psi(t)\rangle$ and $|\psi^\prime(t)\rangle$ in Eq.~\eqref{eq_fidelity} are prepared for the same disorder distribution, the fidelity stays at $F(t)=1$ for $\gamma=0$ if the dynamics starts from an identical initial state. Remarkably, we find that the fidelity in the DMAL regime $\bar F_\mathrm{MBL}(t)$ reveals an anomalous power-law behavior for times $\gamma t \gg 1$ before saturating to the steady-state value as
\begin{align}
\bar{F}_\mathrm{MBL}(t)\propto\left(\frac{1}{\gamma t}\right)^\alpha.
\label{eq_fidelityMBL}
\end{align}
In Fig.~\ref{fig_Fidelity}, we have calculated the dynamics of the fidelity in the DMAL regime, where we have set $h/J=10$ and $\gamma/J=0.05$. We have obtained the exponent $\alpha$ from the power-law fitting of the data (taken in the time interval $200\le\gamma t\le2000$) as $\alpha=0.55, 0.74, 0.93, 1.07$ $(\pm 0.01)$ for $L=8, 10, 12, 14$, respectively. As shown in Fig.~\ref{fig_linearfitting} in the Supplemental Material, the exponent $\alpha$ seems to be proportional to $L$ for small system sizes. On the other hand, as we depart from the DMAL regime, the numerical results demonstrate that this power-law dependence of the fidelity is smeared and finally vanishes in the MMAL regime \cite{Supple}. This is because the time that the power-law decay should end becomes shorter than the typical timescale of the system such as $1/J$.

Thus, we demonstrate that, though both the disorder and the measurement localize the wave function in the steady state, the speed of relaxation dynamics in the DMAL regime is slow enough to exhibit the power-law behavior of the fidelity. We also find that the overlap between random quantum trajectories in the steady state is suppressed by the disorder and the measurement, while the fidelity is made to be zero in the limit $L\to\infty$ in the entire phase diagram \cite{Supple}. These results conclude that the dynamical property in the DMAL regime is distinct from that in the MMAL regime. 

We note that Eq.~\eqref{eq_fidelityMBL} is reminiscent of the logarithmic growth of the von Neumann entropy in the MBL system following the Lindblad master equation \cite{Levi16}. In Ref.~\cite{Levi16}, it is pointed out that the von Neumann entropy of the whole system $S_\mathrm{MBL}^\mathrm{Lind}(t)$ exhibits the logarithmic growth for times $\gamma t\gg1$ before saturation as $E_\mathrm{dis}[S_\mathrm{MBL}^\mathrm{Lind}(t)]\propto \beta\log(\gamma t)$. As the density matrix in the Lindblad equation $\rho_\mathrm{Lind}(t)$ is related to that in the trajectory dynamics $\rho(t)$ as $\rho_\mathrm{Lind}(t)=E[\rho(t)]$, we find that the trajectory average of the squared fidelity in Eq.~\eqref{eq_fidelity} is given by $E[F(t)^2]=\mathrm{Tr}[\rho_\mathrm{Lind}(t)^2]$. Assuming that the second R\'enyi entropy $S_2(t)=-\log\mathrm{Tr}[\rho(t)^2]$ behaves similarly as the von Neumann entropy, we obtain $E_\mathrm{dis}[\log E[F_\mathrm{MBL}(t)^2]]= -E_\mathrm{dis}[{S_2}_{\mathrm{MBL}}^\mathrm{Lind}(t)]\propto\log(1/\gamma t)^\beta$. If $E_\mathrm{dis}$ commutes with $\log$ and the squared fidelity displays similar behavior as the fidelity, we arrive at Eq.~\eqref{eq_fidelityMBL}. We note that $\beta \propto L\sqrt{J/h}$ is reported in Ref.~\cite{Levi16}, and this value is consistent with the fitting data of the exponent $\alpha$ in Eq.~\eqref{eq_fidelityMBL}, which seems to increase as we increase the system size $L$. However, we stress that  $E[F_\mathrm{MBL}(t)]$ cannot exactly reduce to the quantity calculated from the averaged dynamics of $\rho_\mathrm{Lind}(t)$.

\begin{figure}[t]
\includegraphics[width=8.5cm]{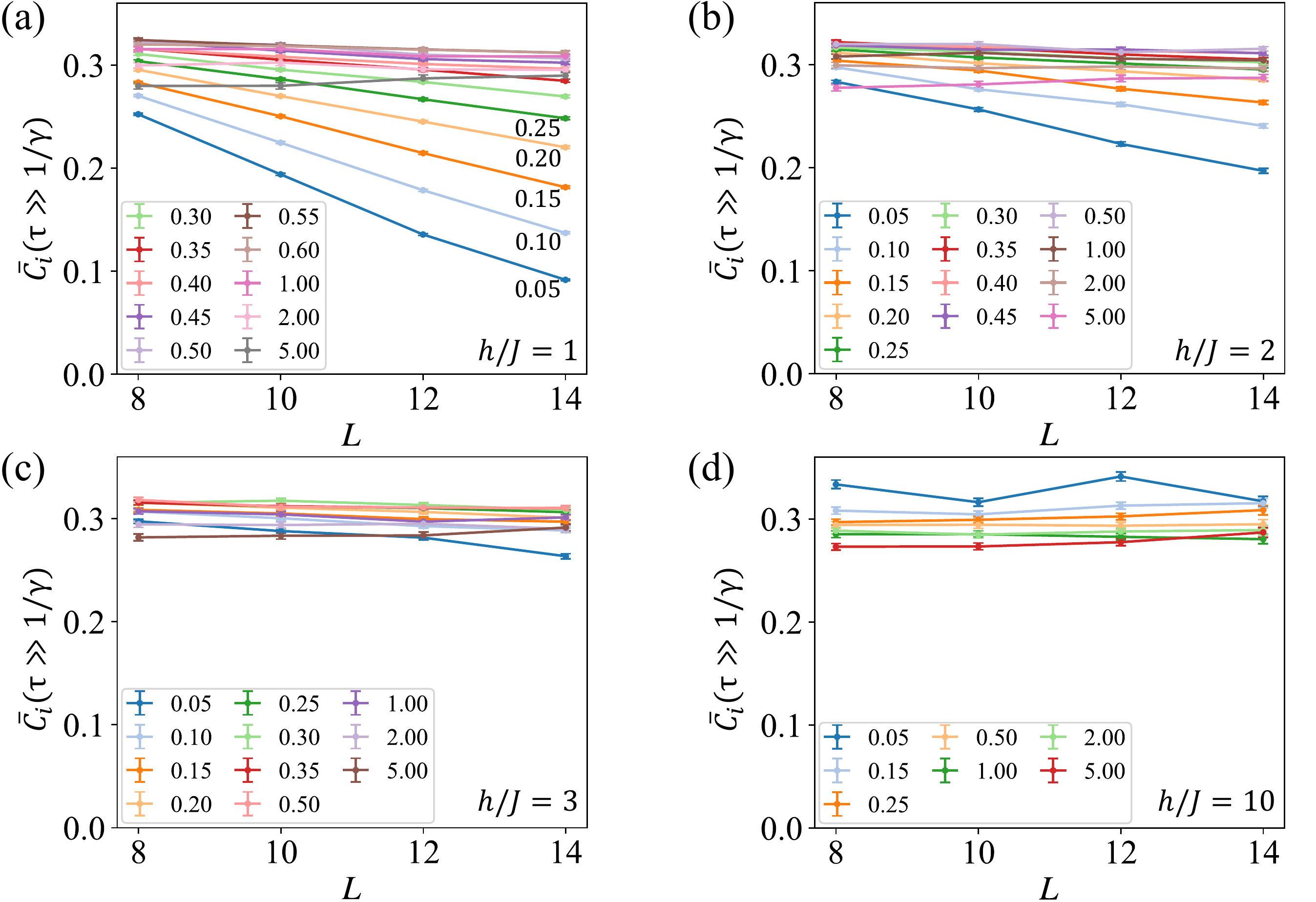}
\caption{System-size dependence of the autocorrelation function $\bar C_i(\tau)$ ($i=L/2+1$) in the long-time regime $\tau\gg1/\gamma$, which takes a disorder-independent value in the MMAL regime and shows slow relaxation reflecting the initial-state information in the DMAL regime. The legend shows $\gamma/J$. The starting point of the autocorrelation is taken at $\gamma t =667$ in (a), (b), and (c), and $\gamma t =2000$ in (d). We set $\gamma \tau=333$ in (a), (b), and (c), and $\gamma \tau =1000$ in (d), and take an average over $300\times100$ realizations for (a) $h/J=1$, (b) $h/J=2$, and (c) $h/J=3$, and $300\times50$ realizations for (d) $h/J=10$.}
\label{fig_AutocorrelationFunction}
\end{figure}

Furthermore, we study the dynamics of the following autocorrelation function:
\begin{align}
C_i(\tau)&=|\lim_{t\to\infty}\langle\psi(t+\tau)| n_i |\psi^{n_i}(t+\tau)\rangle|,
\label{eq_autocorrelation}
\end{align}
where $|\psi^{n_i}(t+\tau)\rangle$ stands for a quantum state that gives the same realization of quantum jump processes as $|\psi(t+\tau)\rangle$, while the particle number operator $n_i$ is inserted into the dynamics at some time $t$ \cite{Supple}. Here, we note that the normalization condition of $|\psi^{n_i}(t+\tau)\rangle$ is given by $||\ket{\psi^{n_i}(t+\tau)}||=||n_i|\psi(t)\rangle||$. The autocorrelation function \eqref{eq_autocorrelation} describes how the particle number operator overlaps with itself under the dynamics from the steady state. Figure \ref{fig_AutocorrelationFunction} shows the autocorrelation function in the long-time regime for the weak disorder case [Figs.~\ref{fig_AutocorrelationFunction}(a), (b), and (c)] and the strong disorder case [Fig.~\ref{fig_AutocorrelationFunction}(d)]. Importantly, we find a characteristic phenomenon that the long-time autocorrelation takes the disorder-independent value around $\bar{C}_i(\tau\gg1/\gamma)\simeq0.3$ in the MMAL regime for the large measurement rate $\gamma/J$. On the other hand, as shown in Fig.~\ref{fig_AutocorrelationFunction}(d), we see that the autocorrelation function in the DMAL regime with weak measurement shows a slow relaxation reflecting the information about the initial N\'eel state (see the plot for $\gamma/J=0.05$). Here, we note that, as we take the site $i$ in $C_i(\tau)$ to be $i=L/2+1$, the dynamics from $t=0$ starts from an up spin for $L=8, 12$, and from a down spin for $L=10, 14$. The difference between the relaxation dynamics means that the initial-state information in the dynamical overlap of the particle number operator is smeared faster in the MMAL regime, where the localization effect caused by the measurement exceeds the one caused by the disorder. Since the dynamics in the DMAL regime is slow enough as shown in the power-law decay of the fidelity in Eq.~\eqref{eq_fidelityMBL}, the slow relaxation of the autocorrelation function agrees with it. In addition, in the volume-law phase [see Figs.~\ref{fig_AutocorrelationFunction}(a), (b), and (c)], we find that $\bar{C}_i(\tau\gg1/\gamma)$ seems to become zero in the limit $L\to\infty$. Since the eigenstate thermalization hypothesis \cite{Rigol08} guarantees $E_\mathrm{dis}[C_i(\tau\to\infty)]\to0.25$ under the unitary dynamics with $\gamma=0$ in the chaotic phase, it is surprising that any finite measurement immediately changes the value of $\bar{C}_i(\tau\gg1/\gamma)$ to zero. To analyze the phenomena in detail, we have further calculated $C_i^\prime(\tau)=|\lim_{t\to\infty}\langle\psi(t+\tau)| \psi^{n_i}(t+\tau)\rangle|$ and found that the behavior of $\bar{C}_i^\prime(\tau\gg1/\gamma)$ is qualitatively similar to that of $\bar{C}_i(\tau\gg1/\gamma)$ (see Supplemental Material \cite{Supple} for detailed results). Thus, we conclude that the behavior shown in Fig.~\ref{fig_AutocorrelationFunction} stems from that of the overlap between the trajectories $|\psi^{n_i}(t+\tau)\rangle$ and $|\psi(t+\tau)\rangle$, though the detailed analysis is left for future work.

\textit{Accessing physical quantities without postselection}.---
Generally, to experimentally realize the trajectory dynamics, we need to postselect special measurement outcomes so as to collect the trajectories that reproduce the same jump processes (see below). However, under continuously monitored dynamics, such experiments require an extremely large number of trials including the factor $(1/(\gamma L\Delta t))^N \times L^N$ for a single quantum trajectory, where $\gamma L \Delta t=:\epsilon \ll1$ represents the accuracy of reproducing the same jump timing and $N$ is the number of quantum jumps. To eliminate this factor, we focus on the fact that $\sum_iL_i^\dag L_i=\sum_i n_i$ is a conserved quantity under the measurement $L_i=n_i$ and propose a general way to realize the trajectory dynamics without postselection \cite{Noel22, Ippoliti21L, Ippoliti22, Fisher22, Nahum22, Dehghani22, Garratt22, Buchhold22} (see Supplemental Material \cite{Supple} for details).

The stochastic dynamics \eqref{eq_stochastic} is constructed from the unitary time evolution described by the Hermitian Hamiltonian \eqref{eq_HB} and the measurement process given by the particle number operator $L_i=n_i$. Then, the evolved state for a single quantum trajectory is written as $|\psi(t)\rangle\propto e^{-iH(t-t_N)}n_{i_N}e^{-iH(t_N-t_{N-1})}n_{i_{N-1}}\cdots e^{-iHt_1}|\psi_0\rangle$, where $\{t_j,i_j\}$ ($j=1, \cdots, N$) is a set of the time and the site of quantum jump processes. Importantly, by using this fact and that the particle number operator is given by $n_i=\sigma_i^z/2+1/2$, where $\sigma_i^z$ is the Pauli matrix, we can decompose the physical quantities into a set of terms including a finite number of $\sigma_i^z$ operators as follows. For example, the fidelity \eqref{eq_fidelity} is constructed from terms such as
\begin{align}
\langle\psi_0|e^{iHt_1}\sigma_{i_1}^ze^{iH(t_2-t_1)}\cdots e^{-iH(t_2^\prime-t_1^\prime)}\sigma_{i_1^\prime}^ze^{-iHt_1^\prime}|\psi_0\rangle,
\label{eq_ExpectationValue}
\end{align}
where $\{t_j^{\prime},i_j^{\prime}\}$ ($j=1, \cdots, N^{\prime}$) is a set of jump processes corresponding to $|\psi^{\prime}(t)\rangle$ and at most $N+N^\prime$ number of $\sigma_i^z$ operators appear. Then, once the jump processes $\{t_j$, $i_j\}$ and $\{t_j^{\prime},i_j^{\prime}\}$ are obtained, we can reproduce the expectation value \eqref{eq_ExpectationValue} by evolving the state under the Hermitian Hamiltonian $H$ and applying a $\sigma^z$ unitary operation in a closed system. We note that, as the R\'enyi entropy has been experimentally observed by using the swap operation \cite{Hastings10, Abanin12, Daley12, Greiner15}, the above method is readily applicable to obtain the entanglement entropy that determines MIPTs. Thus, we can obtain the physical quantities discussed in this Letter without postselection \cite{Noel22, Ippoliti21L, Ippoliti22, Fisher22, Nahum22, Dehghani22, Garratt22, Buchhold22} and can significantly reduce the cost stemming from the factor $(1/\epsilon)^N \times L^N$.

\textit{Conclusion.}---
We have elucidated that dynamical properties of localization induced by the measurement and the disorder are distinct from each other. In particular, in the large-disorder regime with weak measurement, the dynamics of the fidelity is highlighted by the anomalous power-law decay. Our results are readily observed in ultracold atoms, by first combining the off-resonant probe light and the quantum-gas microscopy to realize the continuously monitored dynamics \cite{Ashida16}, and then applying our postselection-free probe method without dissipation. It should be noted that there has been a heuristic argument regarding the existence of the MBL phase in the thermodynamic limit \cite{Footnote}. However, we emphasize that mesoscopic systems such as ultracold atoms in an optical lattice have experimentally demonstrated the existence of chaotic-MBL transitions \cite{Choi16}. Thus, our results are still relevant to experiments regardless of the existence of the MBL phase in the thermodynamic limit. Regarding the finite-size effect, although our calculation is restricted to small system sizes due to numerical limitation, it is important to study the precise system-size dependence of the power-law exponent of the fidelity. Though it is numerically difficult to find critical points in the dynamics of autocorrelation functions, it deserves to study how MIPTs of entanglement are related to them. Moreover, as a similar phase diagram has been obtained in MBL of Liouvillian eigenstates \cite{Hamazaki22}, it is of interest to investigate its relation to MIPTs. Last but not least, since localization properties of monitored free fermions that exhibit Anderson localization have been recently studied \cite{Szyniszewski22, Popperl22, AndersonNote}, it is also interesting how dynamical properties in monitored MBL systems are related to those in monitored Anderson localized systems.

\begin{acknowledgments}
We are grateful to Norio Kawakami for encouraging our research. K.Y. also thanks Masaya Nakagawa for fruitful discussions. The numerical calculations were partly carried out with the help of QuSpin \cite{Weinberg17}. K.Y. was supported by WISE Program, MEXT and JSPS KAKENHI Grant-in-Aid for JSPS fellows Grant No. JP20J21318. We also thank MACS program in Kyoto University for stimulating our collaborations.
\end{acknowledgments}

\nocite{apsrev42Control}
\bibliographystyle{apsrev4-2}
\bibliography{MeasurementMBL.bib}

\begin{thebibliography}{97}
\bibitem [{Foo()}]{Footnote}%
  \BibitemOpen
  \href@noop {} {}\bibinfo {note} {There has been an argument regarding the
  critical point of MBL transitions. For example, in Refs.~\cite{Morningstar22,
  Sels22}, a possibility of a shift of the transition point to a much larger
  value than the one found in numerics is pointed out on the basis of the
  analysis of the avalanche instability \cite{Wojciech17}.}\BibitemShut {Stop}%
\bibitem [{And()}]{AndersonNote}%
  \BibitemOpen
\bibfield  {journal} {  }\href@noop {} {}\bibinfo {note} {In
  Ref.~\cite{Szyniszewski22}, difference of localization properties between
  Anderson localized unitary phase and the measurement-induced area-law regime
  at weak measurement strength has been studied.}\BibitemShut {Stop}%
\end{thebibliography}


\clearpage

\renewcommand{\thesection}{S\arabic{section}}
\renewcommand{\theequation}{S\arabic{equation}}
\setcounter{equation}{0}
\renewcommand{\thefigure}{S\arabic{figure}}
\setcounter{figure}{0}

\onecolumngrid
\appendix
\begin{center}
\large{Supplemental Material for}\\
\textbf{``Localization properties in disordered quantum many-body dynamics \\ under continuous measurement"}
\end{center}

\section{Numerical algorithm for the marked point process}
For practical calculations, we here explain the numerical algorithm for the stochastic Schr\"odinger equation obeying the marked point process \eqref{eq_stochastic} in the main text. The details are summarized in Table~\ref{tab_algorithm}. We note that taking an ensemble average over all possible trajectories reduces to the Lindblad master equation \cite{Daley14}. In our study, the non-Hermitian effective Hamiltonian is rewritten as $H_\mathrm{eff}=H-i\gamma L /4$, and a quantum jump takes place in a time interval $t=-2\log R_1 / (\gamma L)$ depending on a random number $R_1$ (see Table~\ref{tab_algorithm}). This is because the trajectory dynamics under the jump operator $L_i=n_i$ conserves the total particle number of the system. Consequently, the non-Hermitian effective Hamiltonian $H_\mathrm{eff}$ in Table~\ref{tab_algorithm} is replaced by the Hermitian Hamiltonian except for the normalization factor.

\begin{table*}[h]
\caption{Practical algorithm for the quantum trajectory dynamics following the marked point process.}
\centering
\begin{tabular}{l l}\hline \hline
1. & Choose a random number from a uniform distribution $0\le R_1\le1$.\\
2. &Evolve the (initially normalized) state $|\psi(t)\rangle$ until $\langle\psi(t_1)|\psi(t_1)\rangle=R_1$ \\
&according to the nonunitary Schr\"odinger equation $i\partial_t|\psi(t)\rangle=H_{\mathrm{eff}}|\psi(t)\rangle$.\\ 
3. & Make the state $|\psi(t_1)\rangle$ jump with the Lindblad (jump) operator and normalize the state as\\
& $|\psi(t_1+0)\rangle = L_j|\psi(t_1-0)\rangle/\sqrt{\langle\psi(t_1-0)|L_j^\dag L_j|\psi(t_1-0)\rangle}$.\\
& The site $j$ where the jump process occurs is chosen according to the probability\\
& $\langle\psi(t_1-0)|L_j^\dag L_j|\psi(t_1-0)\rangle/\sum_j\langle\psi(t_1-0)|L_j^\dag L_j|\psi(t_1-0)\rangle$.\\
4. &Choose another random number $0\le R_2\le1$ and repeat the processes 2 and 3.\\\hline \hline
\end{tabular}
\label{tab_algorithm}
\end{table*}

\section{Steady-state properties in the MBL phase}
\begin{figure}[h]
\includegraphics[width=13cm]{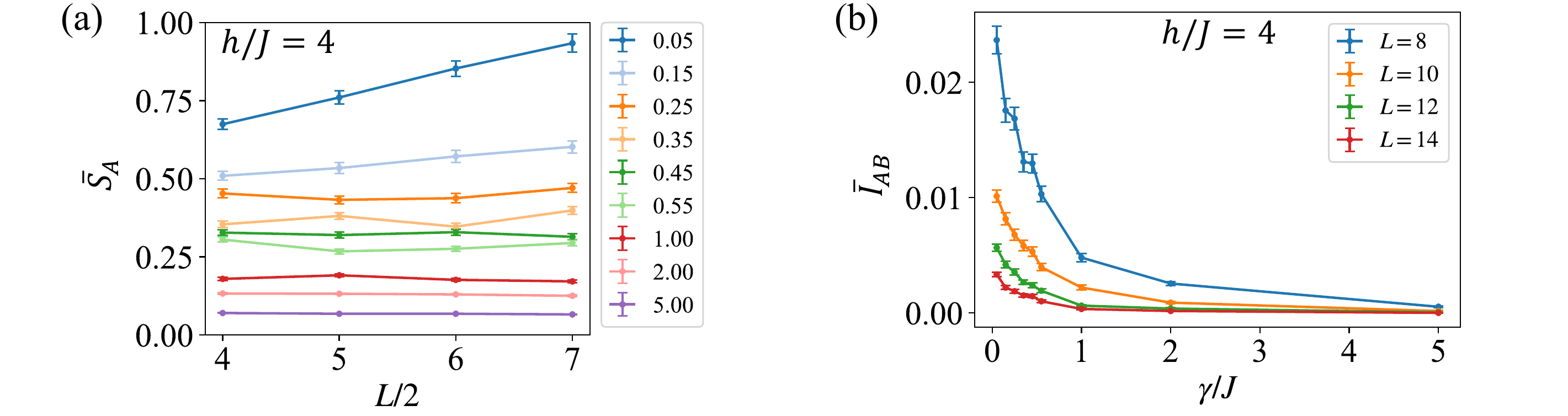}
\caption{(a) Half-chain entanglement entropy $\bar S_A$ ($A=\{j|1\le j\le L/2\}$) with respect to the subsystem size $L/2$ and (b) mutual information $\bar I_{AB}$ ($A=\{1\}$, $B=\{L/2+1\}$) with respect to the measurement rate $\gamma/J$ in the steady state for $h/J=4$. Steady state is reached at $\gamma t=2000$ with the average over $300\times 100$ realizations. The legend in (a) shows measurement rates $\gamma/J$.}
\label{fig_EEMIh4MBL}
\end{figure}

\begin{figure}[h]
\includegraphics[width=13cm]{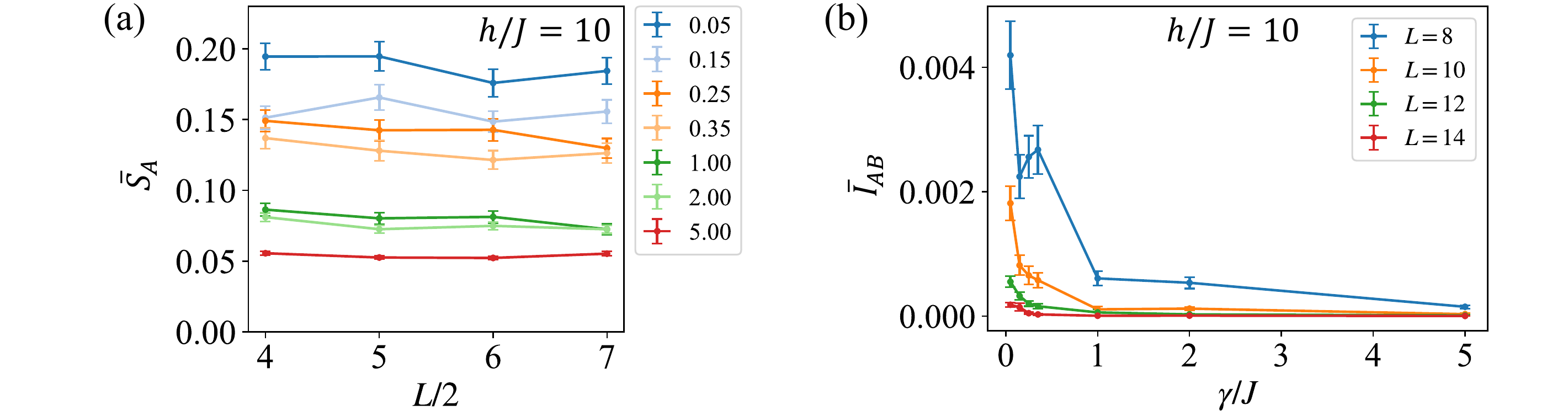}
\caption{(a) Half-chain entanglement entropy $\bar S_A$ ($A=\{j|1\le j\le L/2\}$) with respect to the subsystem size $L/2$ and (b) mutual information $\bar I_{AB}$ ($A=\{1\}$, $B=\{L/2+1\}$) with respect to the measurement rate $\gamma/J$ in the steady state for $h/J=10$. Steady state is reached at $\gamma t=3000$ with the average over $300\times 100$ realizations for $L=8, 10, 12$, and $300\times 50$ realizations for $L=14$. The legend in (a) shows measurement rates $\gamma/J$.}
\label{fig_EEMIDeepMBL}
\end{figure}

In Ref.~\cite{Lunt20}, it was argued that when measurement is introduced to an MBL system with an infinitesimal measurement strength, it leads to an area law of the stationary entanglement. To confirm this, we have calculated the von-Neumann entropy and the mutual information in the steady state in the MBL phase. In Fig.~\ref{fig_EEMIh4MBL}, we have obtained the results for $h/J=4$, which is expected to be just above the critical point of chaotic to MBL transitions in the unitary limit. As the finite-size scaling of $\bar S_A$ shown in Fig.~\ref{fig_EEMIh4MBL}(a) or the peak estimation of $\bar I_{AB}$ shown in Fig.~\ref{fig_EEMIh4MBL}(b) is difficult due to numerical limitation, we have obtained the critical point of MIPTs in the following way. From the results for $h/J=1, 2, 2.5$ for $L=14$, we have found that the mutual information $\bar I_{AB}$ seems to have a peak when the entanglement entropy $\bar S_A$ at $L=14$ takes a value around $20$\% of the maximal entropy $S_\mathrm{max}=\frac{L}{2}\log2$. Thus, we use this value as the threshold to determine the transition point of MIPTs for $h/J=3, 3.5, 4$ (for these values, the peak estimation of $\bar I_{AB}$ is difficult) and estimate that the system already undergoes an area-law transition for $h/J=4$. To further confirm this behavior, we have calculated $\bar S_A$ and $\bar I_{AB}$ in the steady state for $h/J=10$ in Fig.~\ref{fig_EEMIDeepMBL}. In Fig.~\ref{fig_EEMIDeepMBL}(a), we find that the half-chain entanglement entropy in the steady state almost stays constant with respect to $L/2$ within the error bars for any measurement strength. Moreover, we see in Fig.~\ref{fig_EEMIDeepMBL}(b) that the mutual information in the steady state is suppressed to a small value, and there is no peak with respect to the measurement strength $\gamma/J$. According to these results, we conclude that the MBL phase under continuous measurement immediately undergoes an area-law entanglement transition for any finite measurement strength.

\section{Phase diagram for $L=14$}
\begin{figure}[h]
\includegraphics[width=13cm]{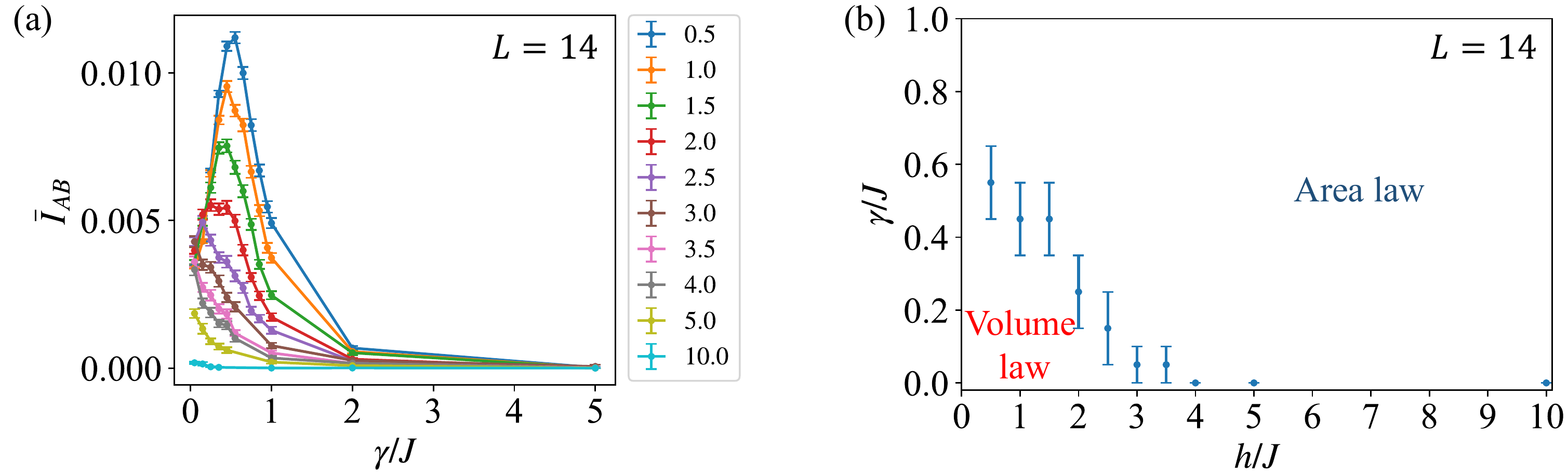}
\caption{(a) Mutual information $\bar I_{AB}$ ($A=\{1\}$, $B=\{L/2+1\}$) in the steady state with respect to the measurement rate $\gamma/J$ and (b) phase diagram, where the phase boundary is obtained from the peak of $\bar I_{AB}$. The legend in (a) shows the disorder strength $h/J$, and the system size is set to $L=14$. Steady state is reached at $\gamma t=1000$ for $0.5\le h/J \le 3$, $\gamma t =1500$ for $h/J=3.5$, $\gamma t = 2000$ for $h/J=4, 5$, and $\gamma t =3000$ for $h/J=10$. We take the average over $300\times 100$ realizations for $0.5\le h/J \le 5$ and $300\times 50$ realizations for $h/J=10$.}
\label{fig_PhaseDiagramL14}
\end{figure}

In Fig.~\ref{fig_PhaseDiagramL14}, we have depicted the mutual information $\bar I_{AB}$ and the phase diagram for $L=14$. From Fig.~\ref{fig_PhaseDiagramL14}(a), we clearly see that the peak of $\bar I_{AB}$ gradually shifts towards lower $\gamma/J$ as we increase the disorder strength, and there is no peak in the MBL phase. From the peak of $\bar I_{AB}$ shown in Fig.\ref{fig_PhaseDiagramL14}(a), we have obtained the phase boundary between the volume-law phase and the area-law phase and have depicted the entanglement phase diagram for $L=14$ as shown in Fig.~\ref{fig_PhaseDiagramL14}(b). Though it is difficult to obtain the exact transition point due to numerical limitation, we estimate that the peak of $\bar I_{AB}$ vanishes around $3\le h/J\le 4$. We note that, as we cannot determine the peak of $\bar I_{AB}$ only from Fig.~\ref{fig_PhaseDiagramL14}(a) for $h/J=3, 3.5, 4$, we have calculated the entanglement entropy with respect to subsystem sizes and have estimated the phase boundary in Fig.~\ref{fig_PhaseDiagramL14}(b) as detailed in the previous section, i.e., we have estimated the transition point of MIPTs when the entanglement entropy $\bar S_A$ at $L=14$ takes a value around $20$\% of the maximal entropy $S_\mathrm{max}=\frac{L}{2}\log2$. It is also noted that the critical measurement rate $\gamma_c(h)/J$ shown in Fig.~\ref{fig_PhaseDiagramL14}(b) monotonically decreases as we increase the disorder strength. From these results, we have obtained the phase diagram Fig.~\ref{fig_PhaseDiagram} in the main text.

\section{Dynamics of the fidelity in the whole phase diagram}
\begin{figure}[h]
\includegraphics[width=15cm]{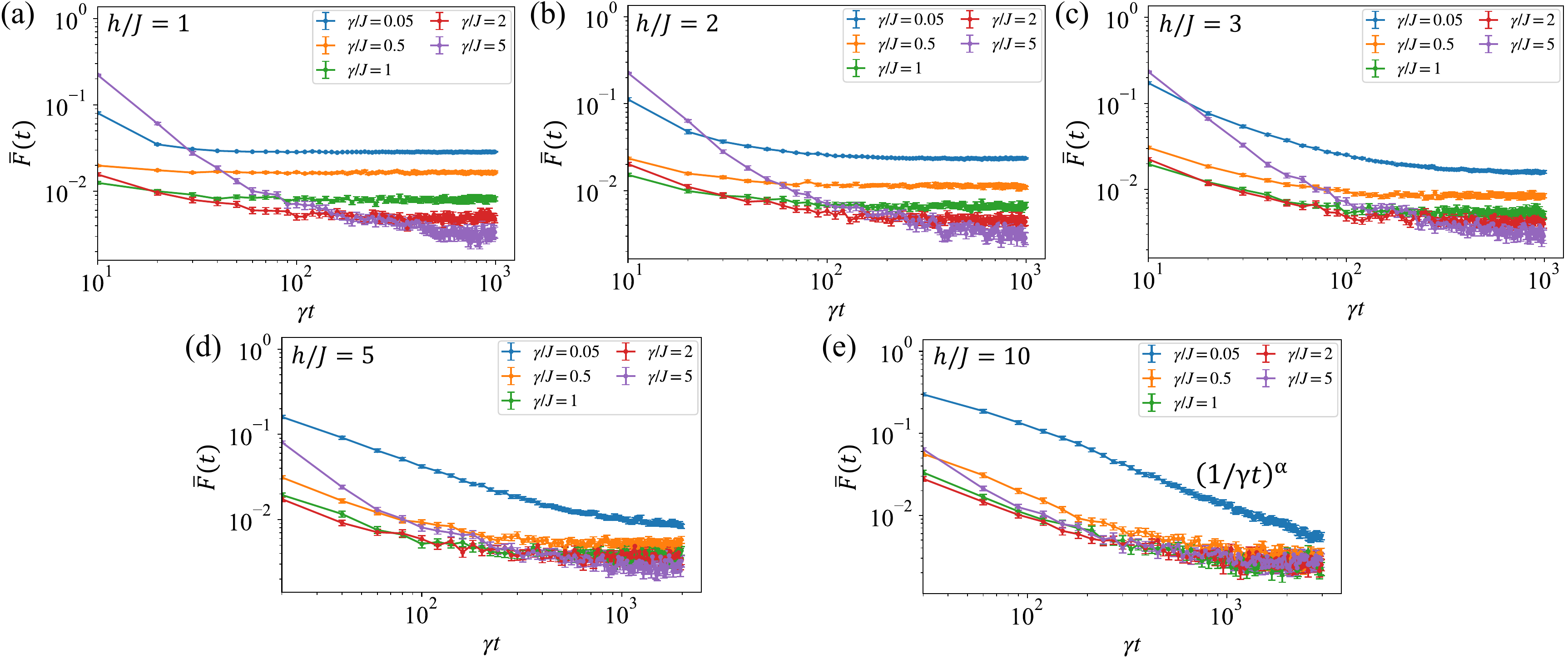}
\caption{Measurement-induced dynamics of the fidelity $\bar F(t)=E_\mathrm{dis}[E[$ $|\langle\psi(t)|\psi^\prime(t)\rangle|$ $]]$ with respect to $\gamma t$ averaged over $100$ disorder realizations for (a) $h/J=1$, (b) $h/J=2$, (c) $h/J=3$, (d) $h/J=5$, and (e) $h/J=10$. For each fixed disorder, we take $100$ pairs of trajectories with respect to $|\psi(t)\rangle$ and $|\psi^\prime(t)\rangle$ for the same initial state. The system size is set to $L=12$.}
\label{fig_FidelityDynamics12}
\end{figure}

\begin{figure}[h]
\includegraphics[width=8cm]{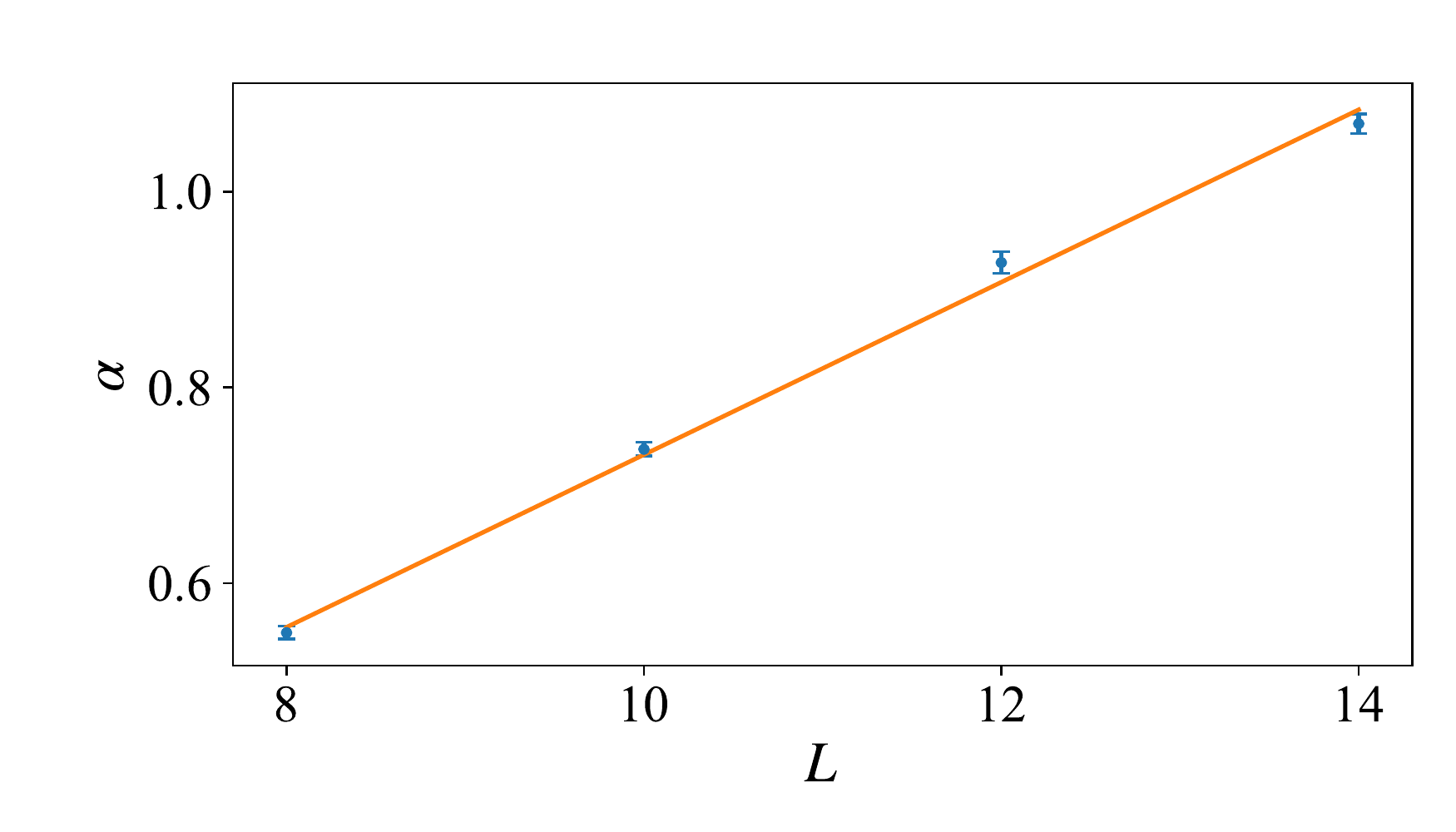}
\caption{Linear fitting of the power-law exponent $\alpha$ of the fidelity in the DMAL regime with respect to the system sizes. We see that $\alpha$ seems to be proportional to $L$ up to the system size $L=14$.}
\label{fig_linearfitting}
\end{figure}

Figure \ref{fig_FidelityDynamics12} shows the dynamics of the fidelity given in Eq.~\eqref{eq_fidelity} in the main text. We change the measurement and the disorder strength and take the average over random disorder and trajectory realizations. First, as seen in Fig.~\ref{fig_FidelityDynamics12}(e) for $h/J=10$ and $\gamma/J=0.05$, we see that $\bar F_\mathrm{MBL}(t)$ exhibits the power-law behavior for about $200\le \gamma t \le 2000$ before saturation. In Fig.~\ref{fig_linearfitting}, we plot the system-size dependence of the power-law exponent $\alpha$ and see that $\alpha$ seems to be proportional to $L$ up to the system size $L=14$. Then, from Figs.~\ref{fig_FidelityDynamics12}(a) to (e), we find that this power-law behavior is gradually smeared and vanishes as the measurement strength is increased or the disorder strength is decreased. Moreover, when we enter the volume-law entanglement phase with weak disorder and weak measurement strength, the decay of the fidelity is faster than the power-law and seems to be an exponential decay to the steady state value [see Figs.~\ref{fig_FidelityDynamics12}(a), (b), and (c)]. Thus, the power-law decay of the fidelity is unique to the DMAL regime (see Fig.~\ref{fig_PhaseDiagram} in the main text), and thus demonstrates that the localization phenomena caused by the disorder and the measurement are dynamically distinct from each other.

\section{Fidelity in the steady state in the whole phase diagram}
\begin{figure}[h]
\includegraphics[width=15cm]{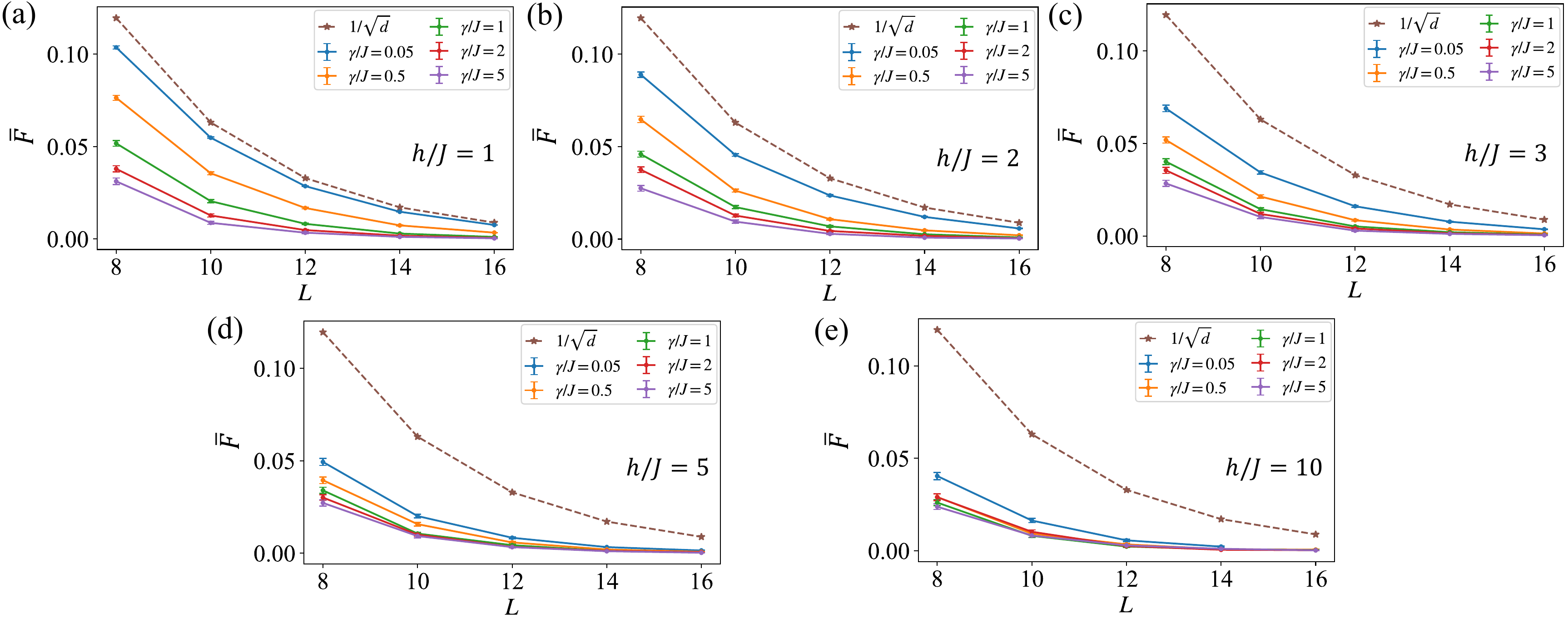}
\caption{System-size dependence of the fidelity $\bar F=E_\mathrm{dis}[E[$ $|\langle\psi|\psi^\prime\rangle|$ $]]$ in the steady state averaged over $100$ disorder realizations for (a) $h/J=1$, (b) $h/J=2$, (c) $h/J=3$, (d) $h/J=5$, and (e) $h/J=10$. For each fixed disorder, we take $100$ pairs of trajectories with respect to $|\psi\rangle$ and $|\psi^\prime\rangle$ for the same initial state. The steady state is reached at $\gamma t =1000$ in (a), (b), and (c), $\gamma t =2000$ in (d), and $\gamma t =3000$ in (e). In all figures, we plot $1/\sqrt{d}$, where $d=L!/((L/2)!)^2$ is the dimension of the Hilbert space for half-filling.}
\label{fig_FidelitySteadyState}
\end{figure}

Figure \ref{fig_FidelitySteadyState} shows the system-size dependence of the fidelity in the steady state. We change the measurement strength and the disorder strength and take the average over random disorder and trajectory realizations. In the entire regime, we see that the fidelity in the steady state approaches to zero in the limit $L\to\infty$. For comparison, in Figs.~\ref{fig_FidelitySteadyState}(a) to (e), we plot $\bar F=1/\sqrt d=1/\sqrt{L!/((L/2)!)^2}$, which is the fidelity calculated from the completely random states at half filling. We find that, as the disorder strength is decreased from Figs.~\ref{fig_FidelitySteadyState}(e) to (a), the fidelity at weak measurement rate $\gamma/J=0.05$ (blue line) gradually approaches to $\bar F=1/\sqrt d$. This result is consistent because the regime with weak disorder and weak measurement corresponds to the volume-law entanglement phase as seen in Fig.~\ref{fig_PhaseDiagram} in the main text, while such a volume law of entanglement entropy also occurs for random states. In the regime with weak disorder [see Figs.~\ref{fig_FidelitySteadyState}(a) to (c)], we clearly see that the fidelity is suppressed as the measurement strength is increased. This means that the measurement localizes the wave function and suppresses the overlap between random quantum trajectories. This result also supports the phase diagram for MIPTs given in Fig.~\ref{fig_PhaseDiagram} in the main text.

\section{Detailed algorithm for calculating the autocorrelation function}
We explain the detailed algorithm for practical calculations of the autocorrelation function given in Eq.~\eqref{eq_autocorrelation} in the main text, which is explicitly written down for a single quantum trajectory as
\begin{align}
C_i(\tau)&=|\lim_{t\to\infty}\langle\psi(t+\tau)| n_i |\psi^{n_i}(t+\tau)\rangle|\notag\\
&=\left|\lim_{t\to\infty}\frac{\langle\psi(t)|e^{iH\tau_1}\cdots n_{i_N}e^{iH(\tau-\tau_N)}n_ie^{-iH(\tau-\tau_N)}n_{i_N}\cdots e^{-iH\tau_1}n_i |\psi(t)\rangle \cdot||n_i|\psi(t)\rangle||}{||e^{-iH(\tau-\tau_N)}n_{i_N}\cdots e^{-iH\tau_1}|\psi(t)\rangle ||\cdot||e^{-iH(\tau-\tau_N)}n_{i_N}\cdots e^{-iH\tau_1}n_i |\psi(t)\rangle||}\right|,
\label{eq_AutocorrelationDetail}
\end{align}
where
\begin{align}
|\psi(t)\rangle= \frac{e^{-iH(t-t_{N^\prime})}n_{i^\prime_{N^\prime}}e^{-iH(t_{N^\prime}-t_{N^\prime-1})}n_{i^\prime_{N^\prime-1}}\cdots e^{-iHt_1}|\psi_0\rangle}{||e^{-iH(t-t_{N^\prime})}n_{i^\prime_{N^\prime}}e^{-iH(t_{N^\prime}-t_{N^\prime-1})}n_{i^\prime_{N^\prime-1}}\cdots e^{-iHt_1}|\psi_0\rangle||}.
\end{align}
Here, we have introduced $\{\tau_j,i_j\}$ ($j=1, \cdots, N$) and $\{t_j,i_j^{\prime}\}$ ($j=1, \cdots, N^{\prime}$), which are sets of the time and the site of quantum jump processes for time duration $\tau$ and $t$, respectively. We remark that, as $n_i^2=n_i$ for hard-core bosons, $C_i(\tau=0)$ is equal to the expectation value of the particle number by definition. Then, the normalization condition of $|\psi^{n_i}(t+\tau)\rangle$ should be $||\ket{\psi^{n_i}(t+\tau)}||=||n_i|\psi(t)\rangle||$. Finally, we give the numerical algorithm to calculate the autocorrelation function \eqref{eq_AutocorrelationDetail} in Talble~\ref{tab_algorithm_auto}. 

\begin{table*}[h]
\caption{Practical algorithm for calculating the autocorrelation function.}
\centering
\begin{tabular}{l l}\hline \hline
1. & Evolve the initial state $|\psi_0\rangle$ by the marked point process and obtain the normalized state $|\psi(t+\tau)\rangle$. \\
& Save the state $|\psi(t)\rangle$ at $\tau=0$ (Here, we set $t\gg1/\gamma$).\\ 
2. & Multiply the state $|\psi(t)\rangle$ by $n_i$ and evolve $n_i|\psi(t)\rangle$ according to the same jump realization as $|\psi(t+\tau)\rangle$ (We define the evolved state \\
& as $|\psi^{n_i}(t+\tau)\rangle$). We note that the normalization condition of $|\psi^{n_i}(t+\tau)\rangle$ is given by $||\ket{\psi^{n_i}(t+\tau)}||=||n_i|\psi(t)\rangle||$.\\
3. &Calculate the autocorrelation function as $C_i(\tau)=|\langle\psi(t+\tau)|n_i|\psi^{n_i}(t+\tau)\rangle|$.\\\hline \hline
\end{tabular}
\label{tab_algorithm_auto}
\end{table*}

\section{Dynamics of the autocorrelation function in the whole phase diagram}
\begin{figure}[h]
\includegraphics[width=15cm]{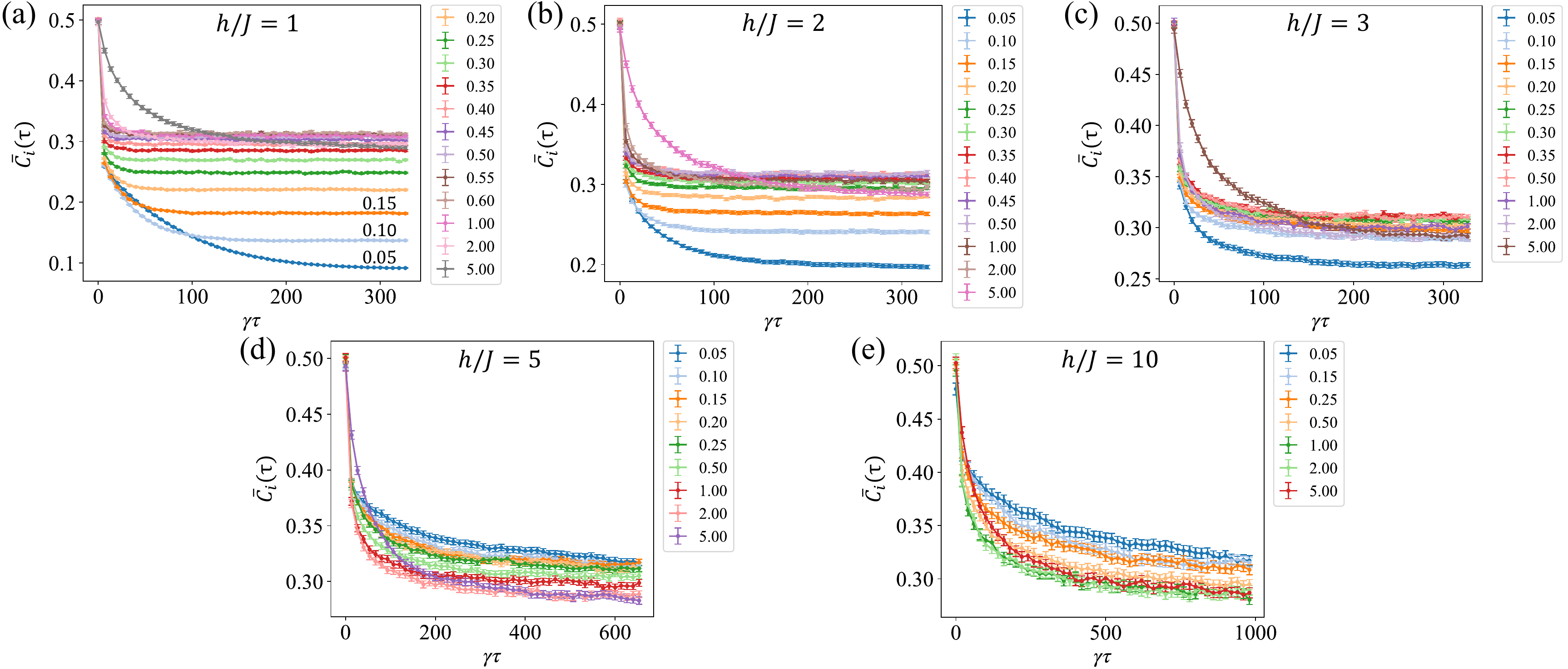}
\caption{Measurement-induced dynamics of the autocorrelation function $\bar C_i(\tau)$ ($i=L/2+1$) with respect to $\gamma \tau$ averaged over $300\times100$ realizations for (a) $h/J=1$, (b) $h/J=2$, (c) $h/J=3$, and (d) $h/J=5$, and averaged over $300\times50$ realizations for (e) $h/J=10$. The parameters are set to $L=14$, and the starting point of the autocorrelation $t\gg1/\gamma$ is set at $\gamma t =667$ in (a), (b), and (c), $\gamma t =1333$ in (d), and $\gamma t =2000$ in (e). The legend shows measurement rates $\gamma/J$.}
\label{fig_AutocorrelationFunctionDynamics14}
\end{figure}

In Fig.~\ref{fig_AutocorrelationFunctionDynamics14}, we plot the dynamics of the autocorrelation function in the entire range of the disorder and the measurement. We take the average over random disorder and trajectory realizations. In the entire range of the disorder, we find that the autocorrelation function approaches almost the same value around $\bar{C}_i(\tau\gg1/\gamma)\simeq0.3$ under strong measurement. On the other hand, from Figs.~\ref{fig_AutocorrelationFunctionDynamics14}(a), (b), and (c), we find that $\bar C_i(\tau)$ with weak measurement and disorder falls below $\bar C_i(\tau\gg1/\gamma)\simeq0.3$. These phenomena may be related to MIPTs obtained in the entanglement phase diagram (see Fig.~\ref{fig_PhaseDiagram} in the main text). 

In contrast, from Figs.~\ref{fig_AutocorrelationFunctionDynamics14}(a) to (e), we see that $\bar C_i(\tau=0)\simeq0.5$ regardless of the disorder strength. As $C_i(\tau=0)$ is equal to the expectation value of the particle number, $\bar C_i(\tau=0)\simeq0.5$ means that the averaged particle number should saturate to that of a thermal state. This is easily understood from the correspondence of the expectation value between the trajectory dynamics and the Lindblad dynamics as follows. The trajectory average $E[\cdot]$ of the particle number is calculated as
\begin{align}
E[\mathrm{Tr}[\rho(t)n_i]]=\mathrm{Tr}[\rho_{\mathrm{Lind}}(t)n_i],
\end{align}
where $\rho(t)$ is the density matrix under the trajectory dynamics, $\rho_\mathrm{Lind}(t)$ is that under the Lindblad dynamics, and we have used the relation $E[\rho(t)]=\rho_\mathrm{Lind}(t)$. As the steady state following the Lindblad equation with the jump operator $L_i=n_i$ is given by a maximally mixed state, we obtain
\begin{align}
\bar C_i(\tau=0)=E_\mathrm{dis}[E[\mathrm{Tr}[\rho(t\to\infty)n_i]]]=E_\mathrm{dis}[\mathrm{Tr}[\rho_{\mathrm{Lind}}(t\to\infty)n_i]]=0.5,
\label{eq_InfiniteTemperature}
\end{align}
which is consistent with the results shown in Fig.~\ref{fig_AutocorrelationFunctionDynamics14}.

\section{Further results on the autocorrelation function}
\begin{figure}[h]
\includegraphics[width=13cm]{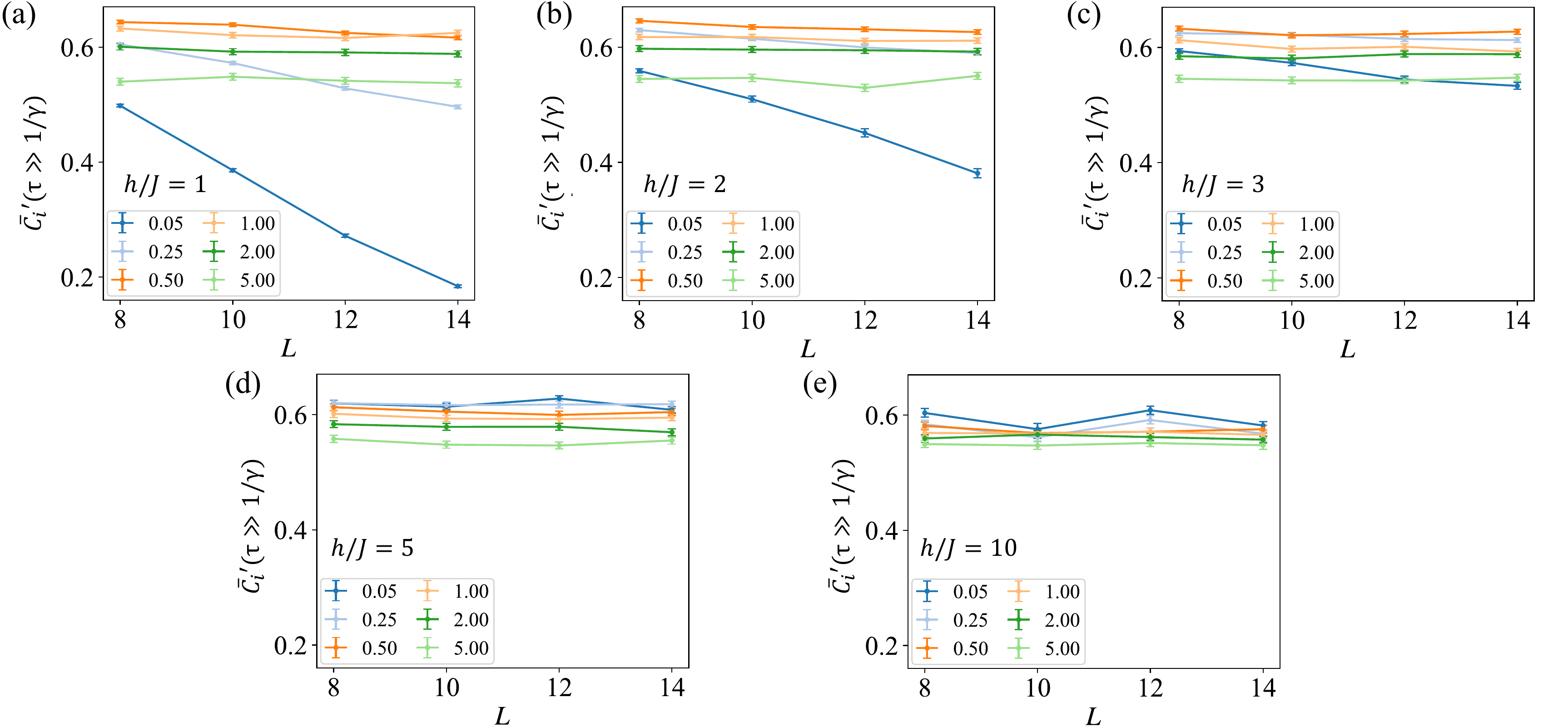}
\caption{System-size dependence of $\bar C_i^\prime(\tau)$ ($i=L/2+1$) in the long-time regime $\tau\gg1/\gamma$ averaged over $100\times100$ realizations for (a) $h/J=1$, (b) $h/J=2$, (c) $h/J=3$, (d) $h/J=5$, and (e) $h/J=10$. The starting point of the dynamics is set to the same value as $\bar C_i(\tau)$ (see Fig.~\ref{fig_AutocorrelationFunction} in the main text) at $\gamma t =667$ in (a), (b), and (c), $\gamma t =1333$ in (d), and $\gamma t =2000$ in (e). We set $\gamma \tau=333$ in (a), (b), and (c), $\gamma \tau =667$ in (d), and $\gamma \tau =1000$ in (e), which are also set to the same value as $\bar C_i(\tau)$ in Fig.~\ref{fig_AutocorrelationFunction} in the main text. The legend shows measurement rates $\gamma/J$.}
\label{fig_Psi_nPsi_overlap}
\end{figure}

To further investigate the behavior of the autocorrelation function \eqref{eq_autocorrelation} in the main text, we calculate the overlap between the states $|\psi^{n_i}(t+\tau)\rangle$ and $|\psi(t+\tau)\rangle$, which is given by 
\begin{align}
C_i^\prime(\tau)=|\lim_{t\to\infty}\langle\psi(t+\tau)| \psi^{n_i}(t+\tau)\rangle|.
\label{eq_Psi_nPsiOverlap}
\end{align}
We note that $\bar C_i^\prime(\tau=0)=0.5$ directly from Eq.~\eqref{eq_InfiniteTemperature}. We also remark that $C_i^\prime(\tau)$ reduces to the mean particle number in a closed system because the Hermitian system is described by the unitary time evolution. In Fig.~\ref{fig_Psi_nPsi_overlap}, we plot the system-size dependence of $\bar C_i^\prime(\tau)$ in the long-time regime $\tau\gg1/\gamma$ averaged over random disorder and trajectory realizations. In the entire regime of the disorder and the measurement, we find that the behavior of $\bar C_i^\prime(\tau\gg1/\gamma)$ is qualitatively similar to that of the autocorrelation function $\bar C_i(\tau\gg1/\gamma)$ (see Fig.~\ref{fig_AutocorrelationFunction} in the main text). Moreover, from Figs.~\ref{fig_Psi_nPsi_overlap} and \ref{fig_AutocorrelationFunction} in the main text, we see that the value of $\bar C_i^\prime(\tau)$ exceeds that of the autocorrelation function $\bar C_i(\tau)$ in the long-time regime $\tau\gg1/\gamma$. Thus, we find that the behavior of the autocorrelation function $\bar C_i(\tau\gg1/\gamma)$ is stemming from the overlap between the trajectories $|\psi^{n_i}(t+\tau)\rangle$ and $|\psi(t+\tau)\rangle$ shown in Fig.~\ref{fig_Psi_nPsi_overlap}, though the quantitative behavior is different from $\bar C_i^\prime(\tau\gg1/\gamma)$.


\section{Estimation the cost of postselection under continuous measurement}
\begin{figure}[h]
\includegraphics[width=12cm]{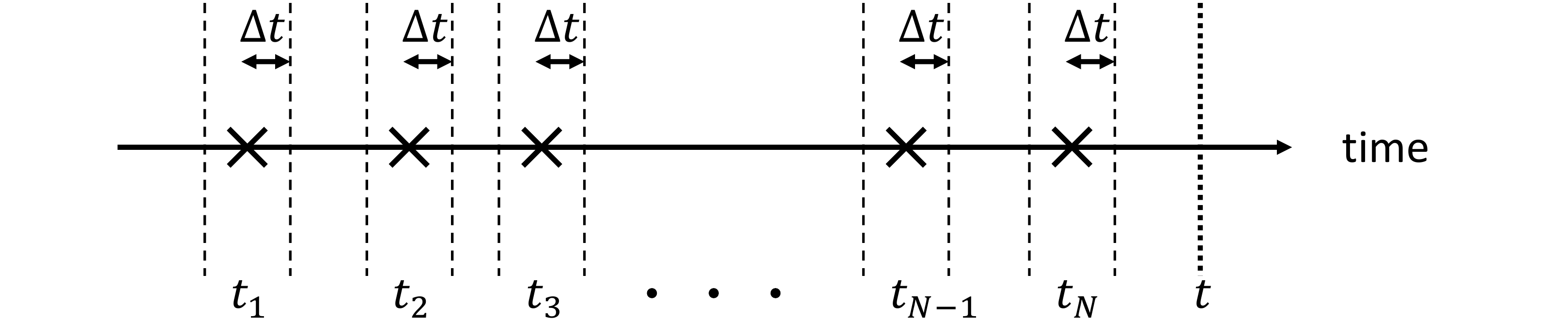}
\caption{Schematic figure of a quantum trajectory dynamics under continuous monitoring, where the positions of quantum jumps are displayed by $N$ cross marks at $t_1,t_2,\cdots,t_N$ along the time axis. We finish the dynamics at time $t$. To postselect the dynamics which reproduces the same jump timing, the positions of quantum jumps should be in the $\Delta t$-neighborhood of $t_i$.}
\label{fig_postselection}
\end{figure}

We here study the cost of postselection to reproduce the continuously monitored dynamics. Figure \ref{fig_postselection}  
shows the schematic illustration of the continuously monitored dynamics for a single quantum trajectory, where jump processes occur at times $t_1,t_2,\cdots,t_N$. To reproduce this dynamics, we have to postselect special outcomes that give the same jump timing. In order to realize such dynamics, we assume that quantum jumps are located in the $\Delta t$-neighborhood of $t_i$ in the reproduced dynamics. Here, $\gamma L \Delta t=:\epsilon \ll 1$ represents the accuracy of reproducing the same jump timing. In our system, as quantum jumps take place in a time interval $-2\log R / (\gamma L)$ depending on a random number $R$ in a uniform distribution $R\in[0,1]$ (see Table \ref{tab_algorithm}), the probability $P_{t_1}$ that gives the first quantum jump in the $\Delta t$-neighborhood of $t_1$ is calculated as
\begin{align}
P_{t_1}=e^{-\frac{\gamma L (t_1-\Delta t)}{2}}-e^{-\frac{\gamma L (t_1+\Delta t)}{2}}\simeq \gamma L \Delta t \cdot e^{-\frac{\gamma L t_1}{2}}.
\end{align}
By repeating a similar process for times $t_2-t_1, t_3-t_2, \cdots, t_N-t_{N-1}$, we obtain the probability $P_{t_N}$ that gives $N$ quantum jumps in the $\Delta t$-neighborhood of $t_1, t_2,\cdots,t_N$ as
\begin{align}
P_{t_N}=(\gamma L \Delta t)^N e^{-\frac{\gamma L t_N}{2}}.
\end{align}
Then, we multiply $P_{t_N}$ by the probability that quantum jumps do not occur in a time interval $[t_N,t]$, and obtain the probability $P_t$ that reproduces the same quantum trajectory as
\begin{align}
P_t=P_{t_N}\cdot e^{-\frac{\gamma L(t-t_N)}{2}}=(\gamma L \Delta t)^N e^{-\frac{\gamma L t}{2}}.
\end{align}
Moreover, the sites where quantum jumps occur should be reproduced at times $t_1, t_2, \cdots t_N$, and such a probability $P_L$ is estimated as 
\begin{align}
P_L\simeq\left(\frac{1}{L}\right)^N,
\end{align}
where we have assumed that the jumps occur randomly at each site with a $N$ equal probability. Thus, to reproduce a single quantum trajectory dynamics, we have to conduct $M_{\mathrm{post}}^{(N)}$ number of trials, where $M_{\mathrm{post}}^{(N)}$ is given by
\begin{align}
M_{\mathrm{post}}^{(N)}=\frac{1}{P_t P_L}=\left(\frac{1}{\epsilon}\right)^N L^N e^{\frac{\gamma L t}{2}},
\end{align}
where $N$ is the order of $\gamma L t$. We note that, to obtain a physical quantity constructed from $|\psi\rangle$ and $|\psi^\prime\rangle$ (corresponding to $N$ and $N^\prime$ quantum jumps, respectively), we need to conduct $M_{\mathrm{post}}^{(N)}+M_{\mathrm{post}}^{(N^\prime)}$ number of trials. To remove the factor $(1/\epsilon)^N\times L^N$, which extremely enhances the cost of experiments, we have proposed a general method to escape postselection in the main text. The cost of this postselection-free probe method is calculated from the number of terms such as Eq.~\eqref{eq_ExpectationValue} in the main text. The number of such terms is estimated as $2^N\times 2^{N^\prime}$, which is directly the experimental cost to reproduce a physical quantity constructed from single quantum trajectories. Thus, we find that the cost stemming from the factor $(1/\epsilon)^N\times L^N$ is removed, and the experimental cost is significantly reduced by our method. We note that, in the DMAL regime where the measurement strength $\gamma$ is small, we can observe the power-law decay of the fidelity for rather small $N$. This is because, though the time taken to reach the steady state becomes long as we decrease $\gamma$, the timescale that the fidelity shows power-law behavior also becomes large, and we can extract the power-law decay for small $\gamma t$.

\end{document}